\documentclass
[preprint,tightenlines,showpacs,prl,onecolumn,letterpaper,runinaddress]{revtex4}%
\usepackage{amsfonts}
\usepackage{amsmath}
\usepackage{amssymb}
\usepackage{graphicx}%
\begin{document}
\title{Train of high-power femtosecond pulses: Probe wave in a gas of prepared atoms}
\author{Gevorg Muradyan$^{a,b}$}
\affiliation{$^{a}$Department of Physics, TU Kaiserslautern, Kaiserslautern, Germany}
\author{Atom Zh. Muradyan$^{b}$}
\affiliation{$^{b}$Department of Physics, Yerevan State University, Yerevan, Armenia}


\begin{abstract}
We present a new method for generating a regular train of ultrashort optical
pulses in a prepared two-level medium. The train develops from incident
monochromatic probe radiation travelling in a medium of atoms, which are in a
quantum mechanical superposition of dressed internal states. In the frame of
linear theory for the probe radiation, the energy of individual pulses is an
exponentially growing function of atom density and of interaction cross
section. Pulse repetition rate is determined by the pump field's generalized
Rabi frequency and can be around 1 THz and greater.

We also show that the terms, extra to the dipole approximation, endow the gas
by a new property: non-saturating dependence of refractive index on dressing
monochromatic field intensity. Contribution of these nonsaturating terms can
be compatible with the main dipole approximation term contribution in the
wavelength region of about ten micrometers (the range of $CO_{2}$ laser) or larger.

\end{abstract}
\maketitle
\tableofcontents

Quantum interference is one of principal values of quantum theory that
essentially enriches the list of earlier ordered principles of physical
theory. Now there is a good comprehension how to prepare the system in a
desirable quantum state of coherent superposition of two or more eigenstates
of a quantum system. This is especially relevant to manipulations with a laser
radiation where some new paths for scientific and technological applications
have been established due to extensive theoretical and experimental studies in
the last two-three decades, e.g. quantum information processing \cite{1},
lasing without inversion \cite{2}.

The scenario discussed in this paper, where a superposition of two dressed
atomic states is coupled to a weak (probe) quasimonochromatic laser radiation
also takes advantage of just this quantum interference effect. \ Of particular
pertinence, to the subject under consideration, is the work of S.E. Harris and
co-workers, who have suggested and used a Raman-type three level interaction
scheme in $D_{2}$-molecular gas to get series of femtosecond pulses \cite{3}.
The applied lasers, driving the molecular Raman transition slightly
off-resonance, cause the molecules to vibrate in unison. Time-varying
refractive index of such a non stationary gas generates ultrabroad spectrum of
Raman sidebands whose Fourier transform may synthesize a train of desired
ultrashort pulses. The usage of quantum interference effects in order to
manipulate the optical properties of gaseous atomic or molecular medium has by
now been established as a useful and powerful method. In particular, by means
of destructive interference of two coupled (Raman) optical transitions, a
resonant, opaque medium can be made transparent for radiation inducing these
transitions \cite{4}. This is the well-known phenomenon of electromagnetically
induced transparency (EIT) \cite{5}, where the vanishing absorption is
accompanied by an enhanced nonlinear response \cite{6}. It sets up the carrier
idea for a number of promising suggestions and applications, such as, slow
light \cite{7} and light stopping \cite{8} in driven atomic systems. Note
also, that the quantum superposition principle takes off any restriction on
the amount of momentum transferred from one energy level to another during a
single-photon absorption/emission process \cite{9}.

In this paper we suggest and analyze a new theoretical model for producing a
train of ultrashort optical pulses. We show that an incident
quasimonochromatic radiation (probe wave) attains a periodic modulation of
intensity accompanied, in general, by amplification, if the medium atoms have
been prepared in a quantum superposition state of two internal states, dressed
by the pump field. For appropriate gas parameters,concentration and resonance
detuning, the probe wave intensity modulates in a form of regular train of
sharp ultrashort pulses. Distance between these pulses is determined only by
the Rabi frequency of the pump field, while duration and intensity of each
pulse are determined by other parameters also, e.g. by the product of atomic
concentration on interaction cross section.

So we consider a gas of two-level atoms (with energy difference $\hslash
\omega_{0}$ between the excited and ground internal atomic bare states
$\left\vert 2\right\rangle $ and $\left\vert 1\right\rangle $) in a far
off-resonance field of monochromatic radiation, regarded as the pump field.
The spin of relevant to optical transition electron and the possible sublevel
structures, as well as the frequency Doppler shifts will not be taken into
account. \ The atomic(internal) state in dressed state basis is given by
\cite{10}:%
\begin{align}
\Psi_{\pm}  &  =N_{\pm}\left(  \left\vert 1\right\rangle -\frac{2\lambda_{\pm
}}{\Omega}\left\vert 2\right\rangle e^{-i\omega_{p}t+ik_{p}z}\right)
\times\label{1}\\
&  \exp\left(  -\frac{i}{\hslash}(E_{1}+\hslash\lambda_{\pm})t\right)
.\nonumber
\end{align}
The normalization coefficient $N_{\pm}=\Omega/\sqrt{2\Omega^{\prime}%
(\Omega^{\prime}\mp\Delta)}$, where $\Omega=2dE_{p}/\hslash$ and
$\Omega^{\prime}=\sqrt{\Delta^{2}+\Omega^{2}\text{ }}$are the normal and
generalized Rabi frequencies in the pump field, $\lambda_{\pm}=-\Delta
/2\pm\Omega^{\prime}/2$ is the high-frequency Stark shift of energy levels for
positive ($+$) or negative ($-$) \ detuning $\Delta=\omega_{p}-\omega_{0}$,
and\ $E_{1}$ is the energy of the bare state $\left\vert 1\right\rangle $. In
fact the detuning $\Delta$ is assumed much larger than the full linewidth of
the optical transition, and thus spontaneous emission and other inelastic
processes are strongly suppressed and can be neglected for timescales
characteristic for observing the later described effects.

Propagation of a weak probe radiation through a system of dressed atoms can be
described by a wave equation%
\begin{equation}
\left(  \overrightarrow{\nabla}^{2}-\frac{1}{c^{2}}\frac{\partial^{2}%
}{\partial t^{2}}\right)  \overrightarrow{A}_{probe}(\overrightarrow
{r},t)=-\frac{4\pi\rho}{c}\frac{\partial}{\partial t}<\widehat{d}>_{probe},
\label{2}%
\end{equation}
for a probe field vector potential $\overrightarrow{A}_{probe}(\overrightarrow
{r},t)=\overrightarrow{A}(\overrightarrow{r},t)\exp[ik\overrightarrow
{r}-i\omega t]+c.c.$ with slowly varying amplitude $\overrightarrow
{A}(\overrightarrow{r},t)$. $\rho$ is the atom number density and
$<\widehat{d}>_{probe}$ is the atomic dipole moment induced by a probe field.
To acquire the latter, one has to find first the atomic state $\left\vert
\Psi(\overrightarrow{r},t)\right\rangle $ in a combined field of dressing and
probe fields, then implement the ordinary quantum mechanical averaging of the
dipole operator by means of this state vector, and later select terms
proportional to the perturbing field $\overrightarrow{A}(\overrightarrow
{r},t)$. It should be added here that the right side of Eq.(2) examines the
case of isolated atoms, or atoms in a rather dilute gas, where collisional
relaxation and dephasing effects can be ignored.

The atomic state vector has the form%
\begin{align}
\left\vert \Psi(\overrightarrow{r},t)\right\rangle  &  =\left(  \alpha
+C_{+}(\overrightarrow{r},t)\right)  \left\vert \Psi_{+}(z,t)\right\rangle
+\label{3}\\
&  \left(  \beta+C_{-}(\overrightarrow{r},t)\right)  \left\vert \Psi
_{-}(z,t)\right\rangle ,\nonumber
\end{align}
and correspondingly%
\begin{gather}
<\widehat{d}>_{probe}=\left(  \alpha^{\ast}C_{+}+\alpha C_{+}^{\ast}\right)
\left\langle \Psi_{+}\right\vert \widehat{\overrightarrow{d}}\left\vert
\Psi_{+}\right\rangle +\label{4}\\
\left(  \alpha^{\ast}C_{-}+\beta C_{+}^{\ast}\right)  \left\langle \Psi
_{+}\right\vert \widehat{\overrightarrow{d}}\left\vert \Psi_{-}\right\rangle
+\nonumber\\
\left(  \alpha C_{-}^{\ast}+\beta^{\ast}C_{+}\right)  \left\langle \Psi
_{-}\right\vert \widehat{\overrightarrow{d}}\left\vert \Psi_{+}\right\rangle
+\nonumber\\
\left(  \beta^{\ast}C_{-}+\beta C_{-}^{\ast}\right)  \left\langle \Psi
_{-}\right\vert \widehat{\overrightarrow{d}}\left\vert \Psi_{-}\right\rangle
.\nonumber
\end{gather}
where the complex numbers $\alpha$ and $\beta$ (we assume them to be known in
following discussion) represent the probability amplitudes of dressed states
and constitute the mentioned superposition state for the atom in a field of
pump laser radiation. \ Such states can be generated, for example, by using
sufficiently short times for raising the pump field intensity from zero to the
atomic state dressing value. This rapid ramp, eliminating the condition of
adiabaticity, creates for the atom a superposition of two dressed states.
Nonvanishing population of both states requires a switching on time, the
inverse value of which is not much smaller than the frequency detuning . For
instance, if \ $\Delta=2\ast10^{11}Hz$, a switching on time of about $10$ psec
is needed to get perceptible percents in the states superposition. Similar
results may be obtained also by means of well-established technique of Stark
switching on of interaction\cite{11}.

The additional terms $C_{+}(\overrightarrow{r},t)$ and $C_{-}(\overrightarrow
{r},t)$ are perturbations due to interaction with probe radiation which are
proportional to the probe field vector potential amplitude $\overrightarrow
{A}_{probe}(\overrightarrow{r},t)$. They have to be determined from the
Schr\H{o}dinger equation yet. Standard calculations yield%
\begin{align}
i\hbar\partial C_{\pm}/\partial t  &  =\alpha\left\langle \Psi_{\pm
}\right\vert \widehat{V}_{probe\text{ }}\left\vert \Psi_{+}\right\rangle
+\label{5}\\
&  \beta\left\langle \Psi_{\pm}\right\vert \widehat{V}_{probe\text{ }%
}\left\vert \Psi_{-}\right\rangle ,\nonumber
\end{align}
where
\begin{gather}
\widehat{V}_{probe}(\overrightarrow{r_{e}},t)=-\frac{e}{c}\overrightarrow
{A}_{probe}(\overrightarrow{r}_{e},t)\widehat{\overrightarrow{p}}-\label{6}\\
\frac{e^{2}}{mc^{2}}\overrightarrow{A}_{p}(z_{e},t)\overrightarrow{A}%
_{probe}(\overrightarrow{r}_{e},t)\nonumber
\end{gather}
is the probe's interaction energy and is obtained from free atom Hamiltonian
after common substitution$\widehat{\overrightarrow{p}}\rightarrow$
$\widehat{\overrightarrow{p}}-(e/c)\overrightarrow{A}_{T}(\overrightarrow
{r},t)$ (here $\overrightarrow{A}_{T}(\overrightarrow{r},t)$ is the total,
pump + probe vector potential) and selection of terms proportional to the weak
probe field vector potential. Note, the ordinary dipole approximation is
represented by the first term in Eq. (6), hence our analysis extends beyond
the dipole approximation for the probe field.

\qquad Calculation of matrix elements in equations (5) supposes integration
over the optical electron coordinate relative to the atomic center of mass
(a.c.m.), and thus we introduced $\overrightarrow{r}_{e}=\overrightarrow
{r}+\overrightarrow{\xi}$, $\overrightarrow{r}$ \ being the a.c.m. radius
vector. Here we suppose, that the bare atomic states \ have no permanent
dipole moment. Due to opposite spatial parities this immediately exterminates
also the bare state nondiagonal matrix elements $\left\langle 1\right\vert
...\left\vert 2\right\rangle \,$, $\left\langle 2\right\vert ...\left\vert
1\right\rangle $ of the out of dipole approximation term of Hamiltonian (6).
\ Insertion of found solutions of Equations (5) into (4) determines the right
hand side of wave equation (2) as an explicit function of system parameters,
proportional to the probe wave vector potential $\overrightarrow{A}%
_{probe}(\overrightarrow{r},t)$. In the next step we apply to this general
form wave equation the approximation of slowly varying coefficients and attain
the reduced wave equation (r.w.e.) for the slowly varying amplitude
$\overrightarrow{A}(\overrightarrow{r},t)$. \ Some of the right-hand side
terms of obtained r.w.e. are responsible for parametric generation processes,
parametric down conversion and 4-wave parametric amplification respectively.
However in this paper these processes will not be considered and we will
restrict ourselves with a more simple non-parametric propagation. In familiar
formulation this approach, as is well known, leads to determination of the
medium refractive index.

The terms which describe the medium optical properties can now be divided into
two groups: one which contains \ a propagation-type exponential factor
$\ \exp(i\overrightarrow{k}\overrightarrow{r}-i\omega t)$ and the second
group, where the frequency in exponential factors is red- or blue-sideband
shifted by a relatively small amount $\Omega^{\prime}=\sqrt{\Delta^{2}%
+\Omega^{2}}$. \ They represent a bias current on incident probe wave
frequency $\omega$ and on two frequencies symmetrically shifted from $\omega$
by the generalized Rabi frequency $\Omega^{\prime}$. \ Existence of these
sideband components temporarily modulates the bias current. The light
travelling through the medium perceives it as a time varying dielectric
constant and becomes modulated in frequency as well as in amplitude. In
following discussion we are analyzing consequences of this modulation.

So, the r.w.e. can be written in the following symbolic form:%

\begin{gather}
\left(  \overrightarrow{n}\frac{\partial}{\partial\overrightarrow{r}}+\frac
{1}{c}\frac{\partial}{\partial t}\right)  A(\overrightarrow{r},t)=\label{7}\\
i\left(  D+LS\exp\left(  -i\Omega^{\prime}\right)  +RS\exp\left(
i\Omega^{\prime}\right)  \right)  A(\overrightarrow{r},t)\nonumber
\end{gather}
where $\overrightarrow{n}=\overrightarrow{k}/k$ is a unit vector in probe
field propagation direction, in general distinct from the pump field
direction. \ $D$, $LS$ and $RS$(together with $A(\overrightarrow{r},t)$) give
the mentioned direct, red- and blue- sideband components of bias current in a
two-level dielectric gas medium. The first one of them is proportional to
population difference $\left\vert \alpha\right\vert ^{2}-\left\vert
\beta\right\vert ^{2}$ and yields the familiar index of refraction (see
Eq.(9)). It has no conceptual relevance to the superposition nature of the
medium, as we would obtain the same result for a gas medium a unit volume of
which contains $\left\vert \alpha\right\vert ^{2}\rho$ atoms in the ground
state and $\left\vert \beta\right\vert ^{2}\rho$ atoms in the excited one.
\ Superposition nature exhibits itself \ only in the red- and blue-sideband
components, which are proportional to the product of the probability
amplitudes $\alpha$ and $\beta$ (see Eq.(10)). \ It is of special importance
for the following results that though $LS$ and $RS$ are real, in general
$LS\neq RS$ and the contribution of red- and blue-sidebands has real and
imaginary parts.

To solve Eq.(7) we assume as a boundary condition that the probe field is
applied perpendicular at the boundary of the medium and is of infinite extent
in transverse direction. Now (taking into account also the explicit
expressions of $D$, $LS$ and $RS$) the seeking solution for the probe wave
vector potential at $t-(\overrightarrow{n}\overrightarrow{r})/c\geq0$ has a
form
\begin{equation}
A_{probe}(\overrightarrow{r},t)=A_{0}F(\overrightarrow{r},t)e^{i\omega
n_{0}(\omega)\overrightarrow{k}\overrightarrow{r}/c\text{ }-i\omega
t}+c.c.\text{,%
$\backslash$%
} \label{8}%
\end{equation}
where $A_{0}$ is the input wave amplitude,%

\begin{gather}
n_{0}(\omega)=1+\frac{\pi\rho}{2\omega^{2}}\left(  \left\vert \alpha
\right\vert ^{2}-\left\vert \beta\right\vert ^{2}\right)  \times\label{9}\\
\left(
\begin{array}
[c]{c}%
\left(  \frac{d^{2}\omega_{0}^{2}\left(  \Omega^{\prime}-\Delta\right)  ^{2}%
}{\hbar\Omega^{\prime^{2}}}+\frac{e^{2}}{m}\frac{\Omega^{2}}{\Omega^{\prime}%
}\right)  \frac{1}{\omega_{p}-\omega+\Omega^{\prime}}-\\
\left(  \frac{d^{2}\omega_{0}^{2}\left(  \Omega^{\prime}+\Delta\right)  ^{2}%
}{\hbar\Omega^{\prime2}}+\frac{e^{2}}{m}\frac{\Omega^{2}}{\Omega^{\prime}%
}\right)  \frac{1}{\omega_{p}-\omega-\Omega^{\prime}}%
\end{array}
\right) \nonumber
\end{gather}
is the ordinary refractive index and the prefactor%
\begin{equation}
F(\overrightarrow{r},t)=\exp\left(  \frac{2\pi\rho d^{2}\omega_{0}^{2}\Omega
}{\hbar\omega\Omega^{\prime3}}\left(  f_{1}-f_{2}\right)  \right)  \label{10}%
\end{equation}
with%
\[
f_{1}=%
\begin{array}
[c]{c}%
\alpha^{\ast}\beta\left(  1-\exp\left(  -i\Omega^{\prime}\left(
\overrightarrow{n}\overrightarrow{r}\right)  /c\right)  \right)  \times\\
\exp\left(  i\Omega^{\prime}t\right)  \left(  \frac{\Omega^{\prime}+\Delta
}{\omega_{p}-\omega}+\frac{\Omega^{\prime}-\Delta}{\omega_{p}-\omega
+\Omega^{\prime}}\right)
\end{array}
\]
and%
\[
f_{2}=%
\begin{array}
[c]{c}%
\alpha\beta^{\ast}\left(  1-\exp\left(  i\Omega^{\prime}\left(
\overrightarrow{n}\overrightarrow{r}\right)  /c\right)  \right)  \times\\
\exp\left(  -i\Omega^{\prime}t\right)  \left(  \frac{\Omega^{\prime}-\Delta
}{\omega_{p}-\omega}+\frac{\Omega^{\prime}+\Delta}{\omega_{p}-\omega
-\Omega^{\prime}}\right)
\end{array}
,
\]
represents the contribution of red- and blue-sidebands, i.e. the superposition
nature of the atomic state.
\begin{figure}
[ptb]
\begin{center}
\includegraphics[
height=1.3889in,
width=3.1981in
]%
{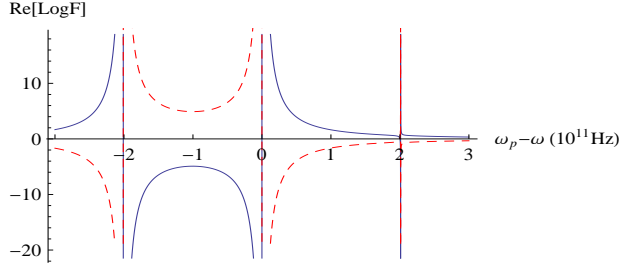}%
\caption{(Color online) The real part of the exponent of prefactor $F$ as a
function of frequency difference of probe and pump fields at $t=\pi
/\Omega^{\prime}$ (solid line) and $t=2\pi/\Omega^{\prime}$(dashed line).
Other parameters are $z=\pi c/\Omega^{\prime}cm$, $\rho=2\cdot10^{15}cm^{-3}%
$,$\alpha=\sqrt{0.99}$, $\beta=0.1$, $\Delta=-2\ast10^{11}Hz$, $\omega
\approx10^{15}Hz$, $\omega_{0}\approx10^{15}Hz$ and $\Omega=2\ast10^{10}Hz$.}%
\end{center}
\end{figure}
\ 

Expression of this prefactor $F(\overrightarrow{r},t)$ is the main result of
this paper. It has imaginary and real parts in the exponent, which mimics the
respective parts of the bias current. The imaginary part of prefactor
$F(\overrightarrow{r},t)$ stipulates a phase modulation, while the real part
introduces amplitude modulation and intensity variance of the probe laser beam
during the propagation in the medium. On the other hand, the exponent is a
periodic function of time and spatial coordinate, which results in a
periodic-type modulation of the probe field intensity, accompanied by
amplification or weakening in average. Fig.1 ascertains the situation in
detail. The solid represents dependence of the real part of exponent on
carrying frequency of probe field at the initial time moment (when the
wavefront reaches the spatial coordinate $z$ examined). Amplification or
weakening of intensity depends on whether the frequency is in a region with
positive or negative values of exponent (of first kind are left and right
regions, of second kind -- the intermediate one). The dashed line pictures the
same dependence for half-period later. Situation is the opposite in this case:
frequencies that were amplified now are decreasing and vice versa. Therefore,
if for a given carrying frequency propagation through the medium starts from
amplification, the probe field in the medium\ is an amplified periodic
sequence of pulses, the minima values of which coincide with the entering
value. If propagation starts from weakening, it gives the opposite regularity,
namely, a sequence of pulses with maxima values repeating the entering value.
\ A periodic flow of energy (photons) takes place between pump and probe
fields. If the flow goes from pump to probe, then one gets the mentioned
amplification (in average) and vice-versa in the opposite case.

Modulation periods $T_{\operatorname{mod}}=2\pi/\Omega^{\prime}$ and
$L_{\operatorname{mod}}=2\pi c/\Omega^{\prime}$ are determined by the
generalized Rabi frequency only, and can be widely tuned and easily controlled
with high precision. \ The depth of modulation, besides the Rabi frequency and
pump and probe frequencies, is determined by the product of gas density on a
single photon scattering cross section. It is evident that deepening of
modulation at a fixed value of repetition period $T_{\operatorname{mod}}$ or
$L_{\operatorname{mod}}$ should result in narrowing of peaks in the modulation
picture and formation of a periodic sequence of sharp (ultrashort) pulses.
Simultaneously choosing the amplification regime, one gets just the desired
comb of high power and ultrashort optical pulses.

The next point which should be brought to attention are zeros of denominators.
In fact, they show processes that have contribution in intensity modulation
(amplification and weakening). \ The condition $\omega=\omega_{p}$ corresponds
to Rayleigh scattering while the other two at $\omega=\omega_{p}\pm
\Omega^{\prime}$ represent common atomic absorption/emission and Stokes
(anti-Stokes) type stimulated hypercombination processes, where absorption
(emission) of two pump photons is followed by emission (absorption) of a probe
photon transferring atom from the lower (upper) energy level to the upper
(lower) one.%
\begin{figure}
[ptb]
\begin{center}
\includegraphics[
height=1.9078in,
width=3.1981in
]%
{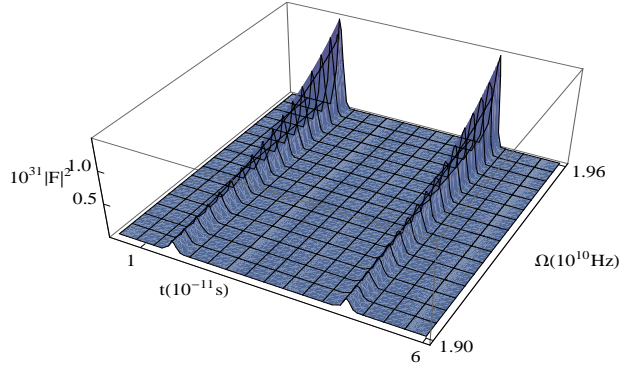}%
\caption{(Color online)Modulation and amplification of the probe wave in
two-level gaseous medium. $\ $Here $\omega_{p}-\omega=$ $2\ast10^{9}Hz,$
$\rho=6\cdot10^{14}cm^{-3}$ and the other parameters are as in Fig.1.}%
\label{Fig.2}%
\end{center}
\end{figure}

And finally, exponent of the prefactor is proportional to the pump intensity
$\Omega$. This means that a simple (linear) superposition of two dressed
quantum states does not make feasible the process of pulse splitting yet. Here
superposition takes place by means of some nonlinear processes, relevant to
the system.

To conceive the picture of probe wave modulation, let's consider the two-level
model of alkali metal gases. \ The characteristic values of $\left\vert
d\right\vert ^{2}$ for dipole allowed transitions are around $2\ast
10^{-34}CGSE$ and sample concentration can be varied in a wide range of
$10^{12}-10^{16}cm^{-3}$. $\ $Typical line broadening is $10^{7}$-$10^{8}$
$Hz$ and therefore \ the lowest allowed (in frame of this model) value for
resonance detuning is $\left\vert \Delta\right\vert =2\pi\times10^{8}Hz$. A
picture of probe wave modulation under some possible conditions is given in
Fig.2. It shows the ability of the presented scheme of interaction to
juxtapose the composition of high repetition ultrashort pulses (duration of an
individual pulse is about $250$ fsec) with their huge amplitude amplification,
desirable for high-field light-matter interactions and implementation of
high-precision comb-spectroscopy, for instance.

When modulation conditions are not satisfied, the probe wave is simply a
running wave propagating in a dielectric medium. Expression (9) for
$n_{0}(\omega)$ (gas refractive index) in comparison with earlier known result
\cite{12} contains out of dipole approximation term $(e^{2}/m)(\Omega
^{2}/\Omega^{\prime})$ in both rectangular brackets. It is interesting that
contribution of this additional term is of resonant nature and has no
saturation in dependence on pump wave intensity $\Omega$. It repeats recently
obtained regularity for the parametric down conversion process \cite{13}. This
out of dipole approximation contribution appears attractive in far IR or
longer wavelength ranges of radiation. For example, in the range of the
$CO_{2\text{ }}$laser operation ( $9.4-10.6$ micrometers) its contribution is
already about $10$ $\%$.

Note that our analytic treatment (in particular, presentation of medium
dispersion properties in the frame of two-level approximation) is essentially
grounded on the usage of slowly varying approximation and implies at least
some tens of femtoseconds for laser light pulse duration

In conclusion, a consistent discussion of the propagation of off-resonance
quasimonochromatic radiation, incident to a gaseous two-level medium prepared
in a quantum superposition state of internal dressed states, shows that it is
split into a sequence of ultrashort pulses with easy and precise tuning of
possible control parameters. The split dynamics can be accompanied, in
general, by large gain or loss in average. Being necessarily based on quantum
nature of superposition in each individual atom, the observation of such
modulation can be regarded as exhibition of a new quantum macroscopic optical phenomenon.

This work was supported by Alexander von Humboldt Foundation, NFSAT/CRDF Grant
UCEP-0702, and by the Armenian Science Ministry Grant 143.

\end{document}